\documentclass[a4paper,10pt]{article}
\usepackage{graphicx,hyperref}
\usepackage{amsmath,amsfonts,amssymb}
\usepackage{fullpage}
\usepackage{authblk}
\usepackage{setspace}
\usepackage{subcaption}
\usepackage{cite}
\usepackage{makecell}

\usepackage{color}

\title{\bf{The Dynamics of Norm Change in the Cultural Evolution of Language}}

\author[a,b]{Roberta Amato}
\author[c]{Lucas Lacasa}
\author[a,b]{Albert D\'iaz-Guilera}
\author[d,\footnote{Corresponding author:  Andrea.Baronchelli.1@city.ac.uk}]{Andrea Baronchelli}
\affil[a]{{\small Departament de F\'isica de la Mat\`eria Condensada, Universitat de Barcelona, Barcelona (Spain)}}
\affil[b]{{\small Universitat de Barcelona Institute of Complex Systems (UBICS), Barcelona (Spain)}}
\affil[c]{{\small School of Mathematical Sciences, Queen Mary University of London, London E1 4NS (UK)}}
\affil[d]{{\small Department of Mathematics, City, University of London, London EC1V 0HB (UK)}}

\date{}
\begin{document}
\maketitle

\begin{abstract}
What happens when a new social convention replaces an old one? While the possible forces favoring norm change - such as institutions or committed activists - have been identified since a long time, little is known about how a population adopts a new convention, due to the difficulties of finding representative data. Here we address this issue by looking at changes occurred to 2,541 orthographic and lexical norms in English and Spanish through the analysis of a large corpora of books published between the years 1800 and 2008. We detect three markedly distinct patterns in the data, depending on whether the behavioral change results from the action of a formal institution, an informal authority or a spontaneous process of unregulated evolution. We propose a simple evolutionary model able to capture all the observed behaviors and we show that it reproduces quantitatively the empirical data. This work identifies general mechanisms of norm change and we anticipate that it will be of interest to researchers investigating the cultural evolution of language and, more broadly, human collective behavior.

\end{abstract}

\section*{Introduction}

Social conventions are the basis for social and economic relations \cite{young2015evolution,ehrlich2005evolution,
bicchieri2005grammar,marmor2009social}. Examples range from driving on the right side of the street, to language, rules of politeness or moral judgments. Broadly speaking, a convention is a pattern of behavior shared throughout a community, and can be defined as the outcome that everyone expects in interactions that allow two or more equivalent actions (e.g., shaking hands or bowing to greet someone) \cite{lewis1969convention,baronchelli2018}. Conventions emerge either thanks to the action of some formal or informal institution, or through a self-organized process in which group level consensus is the unintended consequence of individual efforts to coordinate locally with one another  \cite{ehrlich2005evolution,baronchelli2018}. Crucially, since conforming to a convention is in everyone's best interest when everyone else is conforming too, social conventions are self-enforcing \cite{lewis1969convention}. Yet behaviors change all the time and old conventions are constantly replaced by new ones: words acquire new meanings \cite{croft2000explaining}, orthography evolves \cite{andersen1989understanding}, rules of politeness are updated \cite{watts2005politeness}, and so on.  In isolated groups, shifts in conventions may be driven by the same forces that determine the emergence of a consensus from a disordered state, i.e., institutions or self-organization \cite{croft2000explaining,baronchelli2018}.
However, a quantitative understanding of the processes of norm change has remained elusive so far, probably hindered by the difficulty of accessing adequate empirical data \cite{nyborg2016social}.

 Here, we address this issue by focusing on shifts in orthographic and linguistic norms through the lenses of about $5$ million  
written texts covering the period from 1800 to 2008 from the digitized corpus of Google Ngram \cite{michel2011} dataset. 
Following the same approach that has allowed quantification of processes such as the regularization of English verbs  \cite{lieberman2007quantifying} or the role of random drift in language evolution \cite{Plotkin2017Detecting}, we analyze the statistics of word occurrences for a set of specific linguistic forms that have been historically modified either by language authorities or spontaneously by language speakers in English or Spanish. These include words that have changed their spelling in time and competition between variants of the same word or expression. 
To explore the mechanisms of norm change we consider three separate cases:
\begin{enumerate}
\item \textbf{Regulation by a formal institution.} We analyze the effect of the deliberations of the Royal Spanish Academy, {\it Real Academia Espa\~{n}ola} (RAE), the official royal institution responsible for overseeing the Spanish language, on the spelling of $23$ Spanish words (complete list in SI Sec. 3) \cite{vaquera1986historia,palomo1992,wagner2016,merin2014academia,alcoba2007ortografia,ntlle,casares1954academia}.
\item \textbf{Intervention of informal institutions.} We investigate the effect of dictionary publishing in the US on the updating of American spelling for $900$ words \cite{webdict,ukuslist} (complete list in SI Sec. 10.A).
\item \textbf{Unregulated (or `spontaneous') evolution.} We consider the alternation between forms that are either unregulated or described as equivalent by an institution but have nonetheless exhibited a clear evolutionary trajectory in time (i.e., we do not consider the case of random drift as primary evolutionary force \cite{Plotkin2017Detecting}). In particular, we examine (i) the evolution over time of the use of two equivalent forms for the construction of imperfect subjunctive verbal time in Spanish, for $1,571$ verbs (complete list in SI Sec. 10.C; verbs and declination for each form in \cite{lista_verbi_irr,lista_verbi_reg}),  (ii) the alternation of two written forms of the Spanish adverb {\it solo/s{\'o}lo} (`only') \cite{RAE}, and (iii) $46$ cases of substitution of British forms (e.g., words) with American ones in the US \cite{gonccalves2017fall} (complete list in SI Sec. 10.B).
\end{enumerate}

We show that these mechanisms leave robust and markedly distinct stylized signatures in the data, and we propose a simple evolutionary model able to reproduce quantitatively all of the empirical observations. When a formal institution drives the norm change, the old convention is rapidly abandoned in favor of the new one \cite{osgood1954psycholinguistics,croft2000explaining,blythe2012s,ghanbarnejad2014extracting,lass1997historical}. This determines a universal process of norm adoption which is independent of both word frequency and corpus size.
A qualitatively similar pattern is also observed for norm adoption driven by an informal institution, although in this case the adoption of the new form is smoother and word dependent. In the case of unregulated norm change, the transition from the old to the new norm is slower, potentially occurring over the course of decades, and is often driven by some asymmetry between the two forms, such as the presence of a small fraction of individuals committed to one of the two alternatives \cite{vivian1979spelling,scragg1974history,Martin2014}.

\section*{Data and historical background}

\subsection*{Spanish}
Founded in 1713, the {\it Real Academia Espa\~nola} (Royal Spanish Academy, RAE) is the official institution responsible for overseeing the Spanish language. Its mission is to plan language by applying linguistic prescription in order to promote linguistic unity within and across Spanish-speaking territories, to ensure a common standard in accordance with Article 1 of its founding charter: ``... to ensure the changes that the Spanish language undergoes [...] do not break the essential unity it enjoys throughout the Spanish-speaking world.'' \cite{RAEhistory,RAEstatuto,pochat2001historia}. Its main publications are the {\it Dictionary of Spanish Language} (23 editions between 1780 and today) and its {\it Grammar}, last edited in 2014.
Particularly interesting for our study is the standardization process that the RAE carried out during the 19th century, which enforced the official spelling of a number of linguistic forms \cite{palomo1992,espanola2010ortografia}. 

\begin{figure}
\begin{center}
\includegraphics*[width=0.6\textwidth]{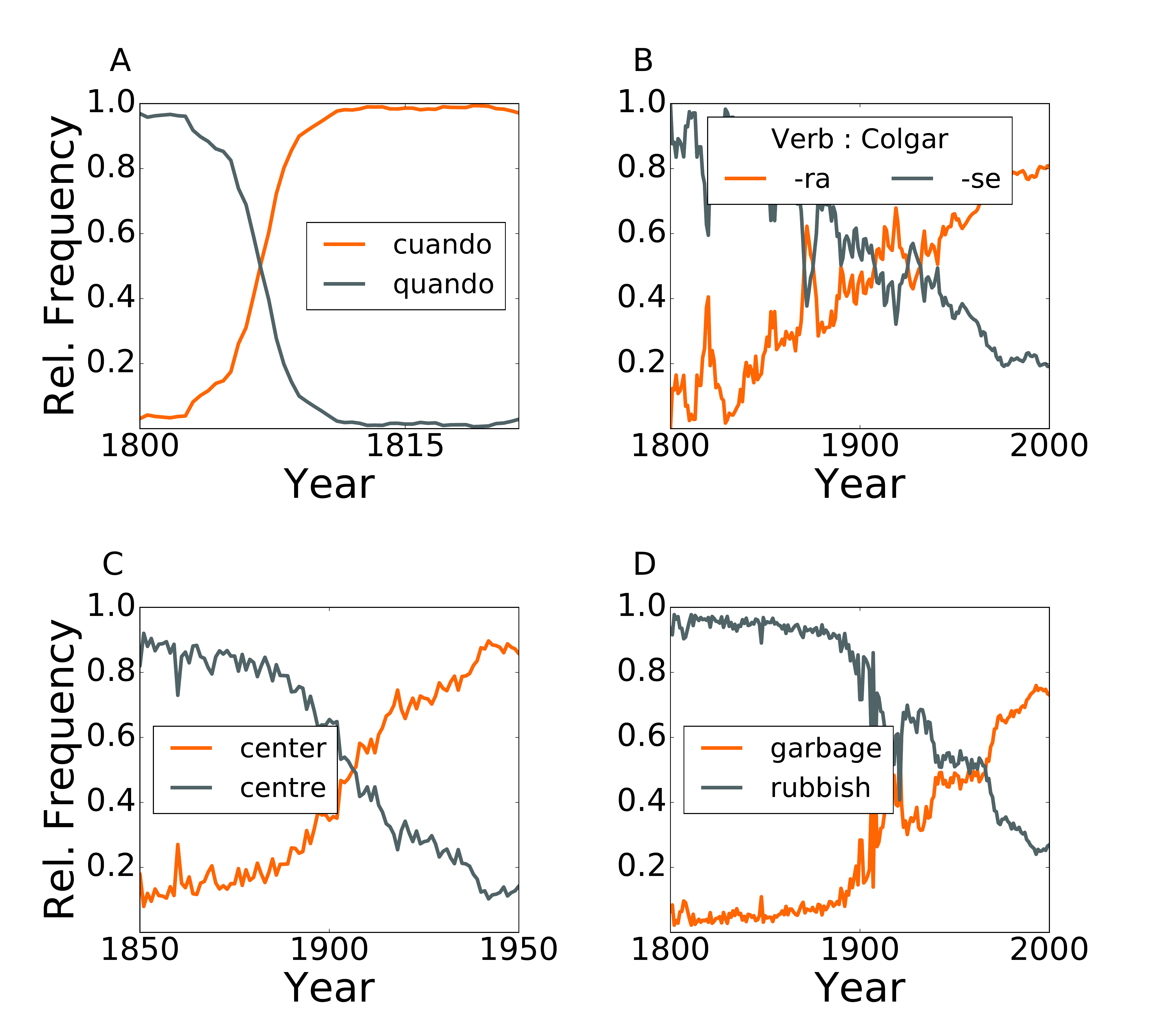}
\caption{Illustrative examples of competing conventions in our dataset (relative frequencies).
(A) Formal institution: the spelling of the Spanish word "quando" (when) was changed into "cuando" by a RAE reform in 1811. (B) Unregulated evolution of two equivalent forms for the past subjunctive, {\it -ra} and {\it -se}, for the verb ``colgar" (to hang). (C) Informal institution: the American Spelling "center" versus the British spelling "centre". (D) Unregulated evolution of ``garbage", the American variant of the British ``rubbish".}
\label{iniziale}
\end{center}
\end{figure}

Our data set contains 23 spelling changes that occurred in four different reforms, in $1815$, $1884$, $1911$ and $1954$ (additional details are in the SI Sec. 3) \cite{vaquera1986historia,palomo1992,wagner2016,merin2014academia,alcoba2007ortografia,ntlle,casares1954academia}.  To illustrate this, Fig. \ref{iniziale}$A$ shows the temporal evolution of the spelling change of the word \textit{quando} (`when') into \textit{cuando} --regulated in the $1815$ reform-- in the Spanish corpus, showing a sharp transition (or ``S-shaped'' behavior \cite{osgood1954psycholinguistics,blythe2012s,ghanbarnejad2014extracting}).
Different is the case of the adverb \textit{solo} (`only'), whose spelling variant \textit{s{\'o}lo} was added to the RAE dictionary in $1956$ after a long unofficial existence supported by a number of academics \cite{RAE,Martin2014,Mellado2017}. We will consider the coexistence of these latter two forms as an example of unregulated evolution (since 2010, the RAE discontinued the \textit{s{\'o}lo} variant again \cite{RAE}, but our dataset does not include such recent data).

A major example of unregulated norm change is offered by the Spanish past subjunctive, which can be constructed in two - equivalent \cite{narajo,kempas2011} - ways  by modifying the verbal root with the (conjugated) ending {\it -ra}  or {\it -se}  (additional details are in SI Sec. 3). For example, the first person of the past subjunctive of the verb \textit{colgar} (`to hang') could be indistinctly \textit{colga-ra} or \textit{colga-se}. 
Fig. \ref{iniziale}$B$ shows the growth of the {\it -ra} variant, for all verbal persons, over two centuries. A similar behavior is found in most Spanish verbs, the form {\it -se} being the most used at the beginning of XIX century (preferred  $\approx 80 \%$ of the times) to the less used at the beginning of the XXI century  (chosen $\approx 20\%$ of the times). This peculiar phenomenon has attracted the attention of researchers for
the last $150$ years and has not been entirely clarified \cite{narajo}. 

Recent results suggest that, whereas individuals typically use only one of the two forms, the alternation between the two variants tends to be found only in speakers who prefer the {\it -se} form \cite{kempas2011,gomez2002}, as also confirmed by a recent analysis of written texts \cite{rosemeyer2017}. Thus, the users of $-ra$ appear to be effectively committed to this unique form. As we will see below, the possibility of such asymmetries of behavior have been incorporated into our model. 

\subsection*{British English vs. American English}
The emergence of American English was encouraged by the initiative of academics, newspapers and politicians -- e.g., US President Theodore Roosevelt \cite{vivian1979spelling} -- who over time introduced and supported new reforms \cite{weinstein1982noah}.
The process gained momentum in the 19th century, when a debate on how to simplify English spelling began in the United States \cite{scragg1974history,venezky1999american,hodges1964short,zachrisson1931four}, which was also influenced by the development of phonetics as a science \cite{wijk1961regularized}. As a result, in 1828 Noah Webster published the first {\it American Dictionary of the English Language}, beginning the Merriam-Webster series of Dictionaries that is still in use nowadays \cite{micklethwait2005noah,scragg1974history}.  Some changes,  such as \textit{color} instead of \textit{colour} or \textit{center} for \textit{centre}, would become the distinctive features of American English. Fig. \ref{iniziale}$C$ shows the transition from the British spelling \textit{centre} to the American \textit{center}. 
 The complete list of the $900$ words examined is reported in SI Sec. 10.A.

The phenomenon of `Americanization' of English \cite{gonccalves2017fall} is not limited to spelling but includes also the introduction of different words or expression which over time replaced the British ones. Recent works \cite{leech2009change,gonccalves2017fall} report how the globalization of American culture might be favoring the affirmation of their specific form of English. We will consider a list of $46$ American-specific expressions in relation to their British counterpart (\cite{gonccalves2017fall}, complete list in SI Sec. 10.B), such as \textit{garbage} vs \textit{rubbish} reported in Fig. \ref{iniziale}$D$, or \textit{biscuit} vs \textit{cookie}.
In all cases, we will consider only books listed in the American English Corpus of Google Ngram. 

\begin{figure}[t]
\begin{center}
\includegraphics*[width=0.5\textwidth]{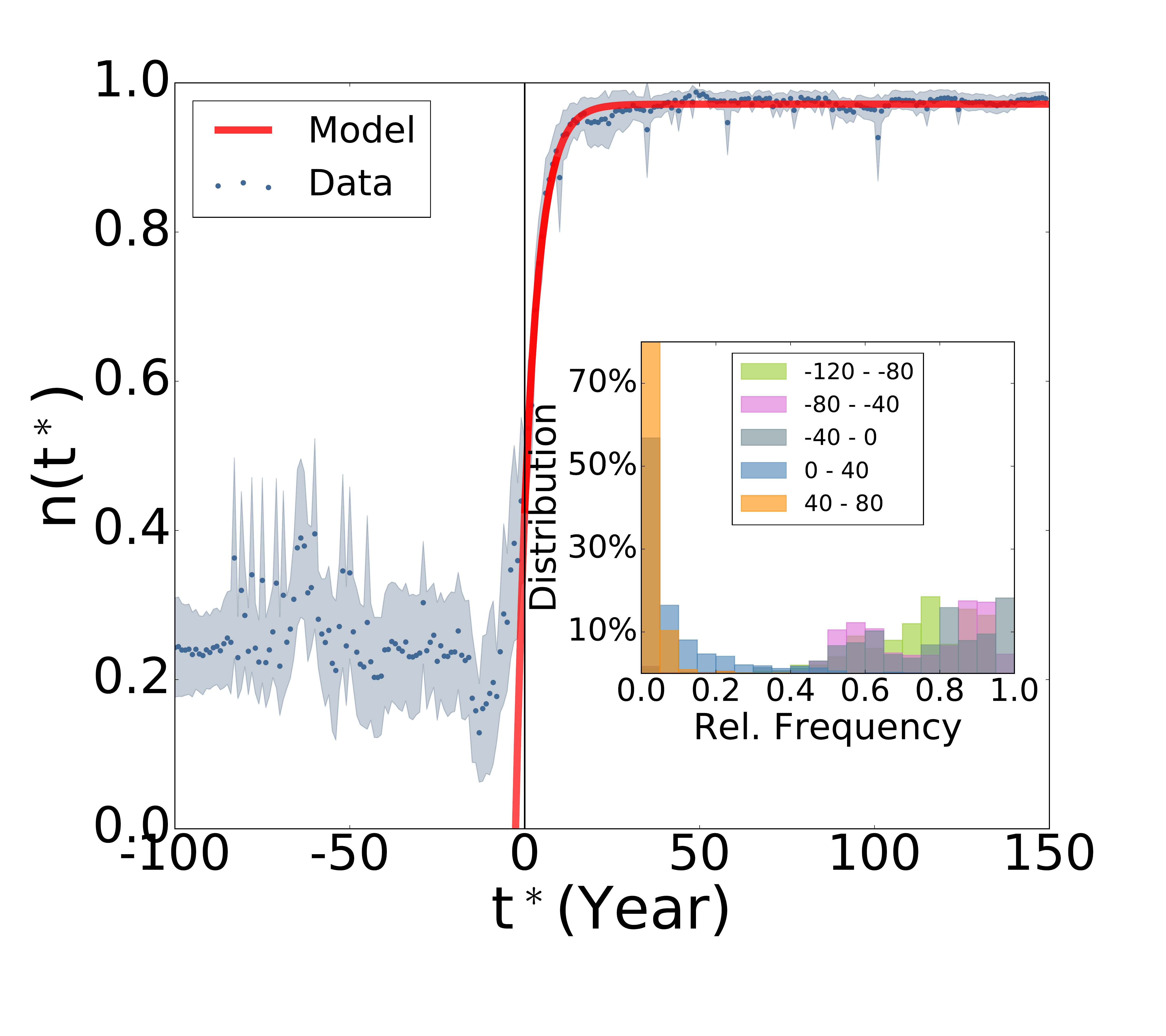}
\caption{Regulation by a formal institution (Spanish, RAE). Main panel: Relative frequency of the new spelling form as a function of the rescaled time  $t^{*}$. Blue points represent the average over all the considered pairs of words and the gray area the standard deviation of the data. The solid line is the prediction of the model outcome (eq. (\ref{eq_sol_a})) after parameter fit ($\chi ^2 = 6 \cdot 10^{-5}$, $ p = 0.99$). The black vertical line denotes the rescaled regulation year $t^{*} = 0$. Inset: Frequency histogram of the old spelling form for all pair of word forms, for different time periods (negative time refers to periods before the regulation). }
\label{rescaling}
\end{center}
\end{figure}

\begin{figure}[t]
\begin{center}
\includegraphics*[width=0.5\textwidth]{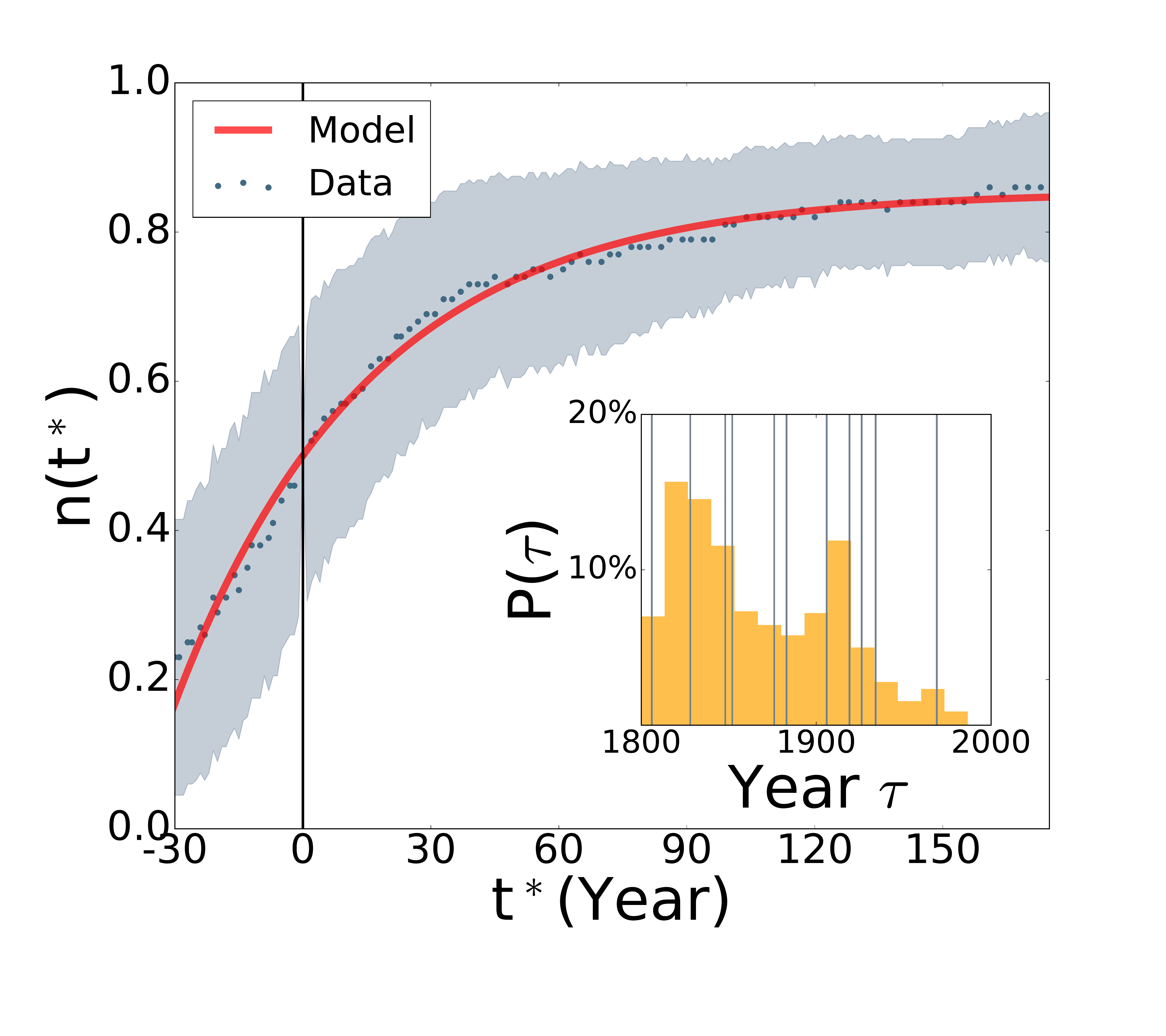}
\caption{Regulation by informal institutions. Main panel: Relative frequency of the American spelling for $900$ English words as a function of the rescaled time $t^{*}$ ($t^{*} = 0$, denotes the surpassing year, for all pairs of words considered). Blue dots represent the average over all the pairs of words and the gray area the standard deviation of the data. The solid line is the model outcome, eq. (\ref{eq_sol_a}), after parameter fit ($\chi ^2 = 8 \cdot 10^{-4}$, $ p = 0.97$). Inset: Distribution $P(\tau)$ of the years $\tau$ in which the American form overcame the British variant for each word. Vertical lines denote important moments of informal regulations of the US spelling such as dictionary editions or spelling updates (additional details are in SI Sec. 5).}
\label{spelling_usa}
\end{center}
\end{figure}

\section*{Model}

We introduce a simple model that describes the evolution in time of two alternative forms of a word (i.e., two alternative conventions). For example, the two norms may represent two spelling alternatives (\textit{-or} vs \textit{-our} as in \textit{color/colour}), two ways to form a verbal tense (\textit{-ra} vs \textit{-se}) or two different words to refer to the same concept (\textit{biscuit} vs \textit{cookie}). 

The model describes a system of books where instances of the two conventions are added by authors through the publication of books. Authors select which convention to use (i.e., which form to introduce in the system) either by following the indications of an institution or considering the current state of language. In the first case, authors simply adopt the recommended norm (or `new norm', for simplicity, as we focus on cases of norm change). In the latter case, the convention to be used is selected with a probability proportional to its current frequency, as in the neutral model for evolution \cite{kimura1983neutral}.  Additionally, some authors can be committed to one specific form, thus being indifferent to any external influence, as suggested by the literature on the study of orthographic norm change in both English \cite{vivian1979spelling,scragg1974history} and Spanish \cite{Martin2014}. When an authority is present, the presence of commitment is revealed by the (empirically verified, see SI Sec. 4 and Fig S1) persistence in time of the old norm, and translates into the model such phenomena as, for example, the re-editions of past books whose orthography is not updated \cite{pechenick2015characterizing}.  In the case of unregulated evolution, committed authors privilege the initially less popular new norm, contributing to its success.

The two different conventions are labeled as `new' and `old', and their number is $\cal N$ and $\cal O$ respectively and the total number of conventions at time $t$ is given by ${\cal W}(t)={\cal N}(t)+{\cal O}(t)$. For a more transparent comparison with the data, aggregated on a yearly basis, we adopt a discrete-time formulation of the model where one time step corresponds to one year. 
The evolution of the densities $n(t)={\cal N}(t)/{\cal W}(t)$ and $o(t) ={\cal O}(t)/{\cal W}(t)= 1-n(t)$ 
is described by the following equations
\begin{eqnarray}
n(t+1) &=&  \left(1-c\right)\left(1-\gamma\right)n(t)+\left(1-c\right)\gamma E_{{\cal N}} +c\,, \nonumber \\
o(t+1) &=&  \left(1-c\right)\left(1-\gamma\right)o(t)+\left(1-c\right)\gamma E_{{\cal O}}\,.
\label{eq_dens}
\end{eqnarray}

New words are inserted by writers (authors). A writer is committed to the use of one specific convention, with probability $c$, or neutral, with probability $1-c$. Neutral writers follow the institutional enforcement, with probability $\gamma$, or sample the current distribution of norms, with probability $1-\gamma$. For simplicity, we assume that each writer inserts just one convention and that the probabilities $c$ and $\gamma$ are constant. When an institution promotes the norm ${\cal N} $, it makes an effort $E_{{\cal N}}  = 1$ and $E_{{\cal O}}=0$ otherwise. If the institution is impartial, both forms are a priori equivalent and $E_{{\cal N}} = E_{{\cal O}} = \frac{1}{2}$. Again for simplicity, in the equations all committed writers privilege the same convention \cite{vivian1979spelling,scragg1974history,Martin2014}.This is the new norm in the above equations, while expressions for the symmetric case of committed agents that support the old form are reported in SI Sec. 1. The general solution of the system of equations (\ref{eq_dens}) is:

\begin{equation}
n(t) = \frac{B}{(1-A)}\left(1-A^{t}\right)+n_{0}A^{t}\,,
\label{eq_sol_a}
\end{equation}

\begin{figure*}
\begin{center}
\includegraphics*[width=1\textwidth]{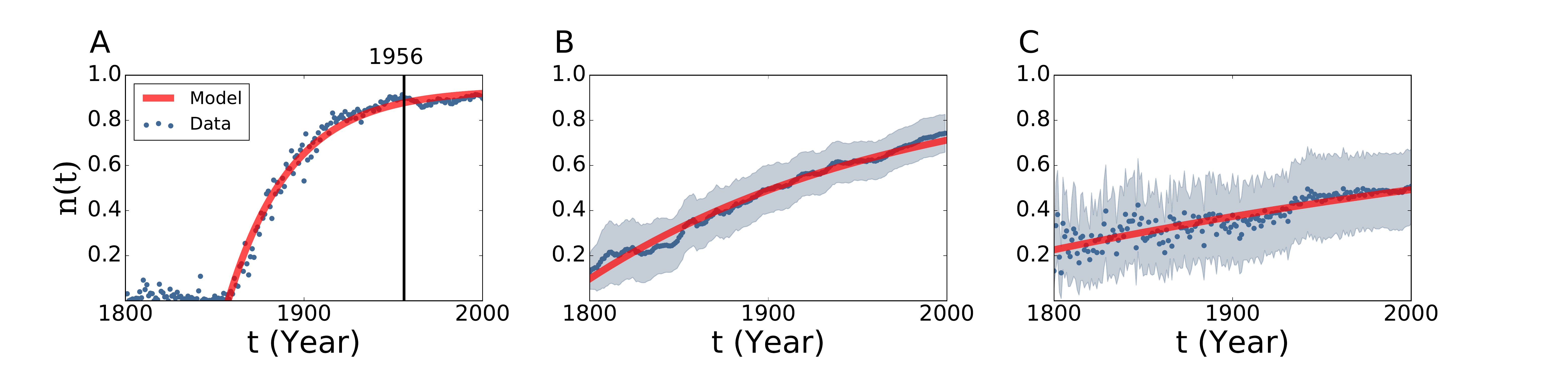}
\caption{Unregulated norm change. (A) Case \textit{s{\'o}lo} versus \textit{solo}. Blue dots represent the relative frequency of the Spanish adverb  \textit{s{\'o}lo} (increasing in detriment of the alternative form \textit{solo}). The solid line is the prediction of eq. (\ref{eq_sol_a}) for this case, after parameter fit ($\chi ^2 = 4 \cdot 10^{-4}$, $ p = 0.98$). The vertical line signs the year $1956$ when RAE intervened explicitly on the case \cite{RAE,Martin2014,Mellado2017}. The curve saturates to a value smaller than $1$ probably due the presence of a percentage of adjectives, indistinguishable from the adverb in the data.
(B) Case of '{\it -ra}' versus '{\it -se}' in Spanish subjunctive. Blue dots represent the relative frequency of the form {\it -ra} (increasing in detriment of the alternative but equivalent form {\it -se}) in Spanish past subjunctive conjugation of verbs, averaged over all verbs considered. Solid line is the specific prediction of eq. (\ref{eq_sol_a}) for this case, after parameter fit ($\chi ^2 = 2 \cdot 10^{-3}$, $ p = 0.96$).
(C) Case of Americanization of English in US. Blue dots represent the relative frequency of the American variant (with respect to the British variant) in US corpus, averaged over all the expressions examined.
Solid line is the specific prediction of eq. (\ref{eq_sol_a}) for this case, after parameter fit ($\chi ^2 = 6 \cdot 10^{-3}$, $ p = 0.93$). For all the cases the grey area identifies the standard deviation of the data.}
\label{spont}
\end{center}
\end{figure*}

\noindent where $A=\left(1-c\right)\left(1-\gamma\right)$, $B=\left(1-c\right)\gamma E_{{\cal N}}+c$ and $n_{0} = n(t = 0)$ (see SI Sec. 2).  
It is worth noticing that, when $E_{{\cal N}}=1$, for $\gamma=1$ eq. (\ref{eq_sol_a}) describes an instantaneous transition in which the new norm immediately saturates to $B/(1-A)$ (with $B/(1-A)=1$ if the commitment supports the new norm, as here, or $B/(1-A)=1-c$ if it supports the old norm, see SI). In this sense, values $\gamma<1$ correspond to a situation in which the response of the system to an institutional intervention is not immediate. 
In the following sections we show that, by appropriately varying the parameter values, the analytic solution Eq. (\ref{eq_sol_a}) reproduces all the empirical observations.

\section*{Results}
\subsection*{Regulation by a formal institution}

In the main panel of Fig. \ref{rescaling} we consider the relative frequency, $n(t)$, of appearance of the new spelling for the 23 words in our dataset affected by RAE reforms \cite{vaquera1986historia,palomo1992,wagner2016,merin2014academia,
alcoba2007ortografia,ntlle,casares1954academia} (See SI Sec. 4). 
By a simple rescaling (translation) of the time axis as $t^{*} = t-t_{r}$ (where $t_{r}$ is the regulation year for each specific pair of conventions), we find that all the experimental curves collapse. The regulatory intervention ($t^{*}=0$) determines an abrupt transition towards the adoption of the new norm. 
This discontinuity is captured by the distribution of the old spelling among the words before and after the regulation in the inset panel of Fig. \ref{rescaling}. Importantly, such rescaling  indicates that the transition is size-independent. For example, for the $1815$ regulation, our dataset consists of $B_{1815}=59$ books and $S_{1815}= 4,149,151$ words, whereas for the regulation enforced in $1954$ we have $B_{1954}= 2774$ books and $S_{1954}= 244,138,299$ words, but transition between the old and new form occurs over approximately the same amount of time in the two cases. 
Model parameters for the case of formal regulation are $E_{{\cal N}}=1$ and commitment supporting the old convention for which $B=\left(1-c\right)\gamma$ (eq. (1) of SI Sec. 1). 
The main panel of Fig. \ref{rescaling} shows that the fit of eq. (\ref{eq_sol_a}) matches the empirical data ($\gamma = 0.2$, $c = 0.006 $ and $n_{0} = 0.42$ from the data). As we will see below, different values of $\gamma$ correspond to different roles played by institutions in the process of norm change (see SI Sec. 7, Fig. S3$A$ and Fig S4$A$ for the behavior of individual curves, and SI Sec. 8 Fig.S4$B$ for the corresponding distribution of $\gamma$). 

\subsection*{Intervention of informal institutions}
\label{sec:2}

We now focus on the dynamics occurring between American and British spelling through the analysis of 900 words as they appear in our US corpus (complete list in SI Sec. 10.A). 
As in the case of formal institution, we have $E_{{\cal N}} = 1$ and the commitment supporting the old (i.e., the British, here) spelling. For each pair of conventions we identify the year $\tau$ in which the British form was
surpassed in popularity by the American one (the inset panel of Fig. \ref{spelling_usa} shows the empirical frequency distribution of these surpassing times $P(\tau)$). The main panel of Fig. \ref{spelling_usa} shows that by rescaling time via simple translation $t^{*}=t-\tau$ all experimental curves collapse, similarly to the above case of formal institution. The model eq. (\ref{eq_sol_a}) reproduces the data. The value of $\gamma=0.02$ obtained in this case is much smaller than the one relative to the above case of formal institution ($\gamma=0.2$), quantifying the weaker role played by informal institutions (other parameters $c=0.003$ obtained by the fit, and $n_0 = n(t^*=0)= 0.5$ by construction). This result is confirmed by analyzing each pair of competing conventions in isolation (see SI Sec. 7, Fig. S3$B$ and Sec. 8, Fig.S4).

\subsection*{Unregulated evolution}

As a third case we explore the process of unregulated norm change by considering the relative frequency of appearance of the form {\it s{\'o}lo} (vs {\it solo}) (Spanish for `only')\cite{RAE} in the Spanish corpus, the relative frequency of appearance in the Spanish corpus of the past subjunctive form ending in $-ra$ and the one ending in $-se$ for $1,571$ verbs  (See SI Sec. 10.C), and the relative frequency of appearance in the US corpus of $46$ cases, among words and expressions, of substitution of British forms for American ones  (See SI Sec. 10.B). Since the institution is impartial, we have $E_{{\cal N}} = E_{{\cal O}} = \frac{1}{2}$.
Figs. \ref{spont}$A$, $B$ and $C$ show that the solution Eq. (\ref{eq_sol_a}) describes well the data relative to growth of the form {\it s{\'o}lo} ($\gamma=3.10^{-3}$, $c=0.02$), the growing of the  $-ra$ form for the subjunctive of Spanish verbs ($\gamma=10^{-17}$, $c=0.005$) and the growth of American forms ($\gamma=10^{-14}$, $c=0.002$), respectively (solid lines correspond to the model predictions after parameter fitting).  The values of $\gamma$ obtained here are significantly smaller than the ones observed for the cases of formal and informal institutions, and corroborate the fact that centralized authorities played essentially no role in this case (see also Fig. S4 in the SI Sec. 8 for the analysis of individual curves). 
It is worth noting that \textit{solo} (without accent) can be also used as adjective and that, while the competition \textit{solo/s{\'o}lo} concerns only the adverb, the data do not allow us to distinguish between the adverb or adjective use. Our analysis shows that the adverb is dominant, as the adverb-specific \textit{s{\'o}lo} is nowadays the most used form, but the non-saturation of the curve in Fig. \ref{spont}$A$ can be interpreted as a signature of the presence of a percentage of adjectives in our dataset.

\begin{figure}[t]
\begin{center}
\includegraphics*[width=0.6\textwidth]{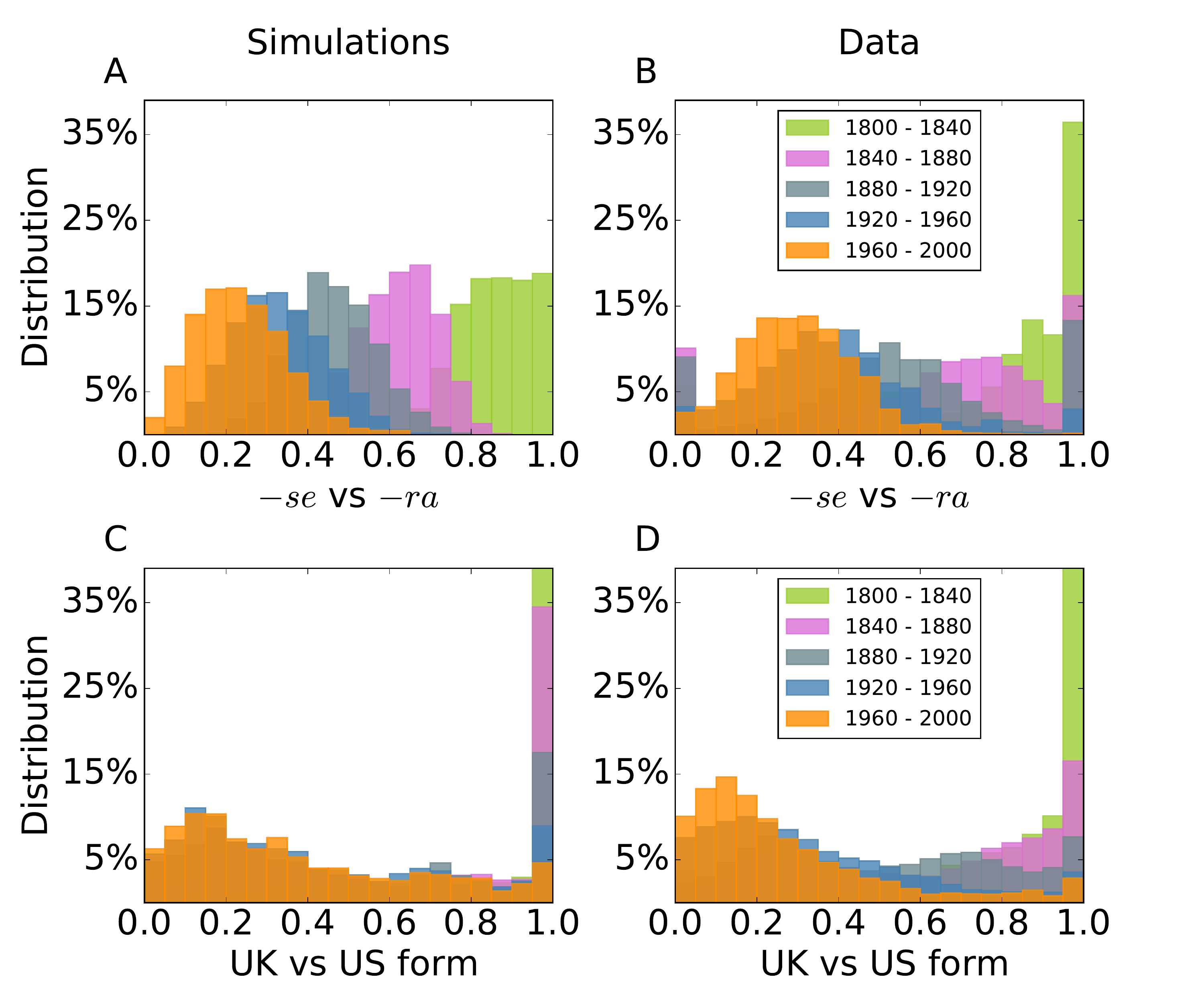}
\caption{Empirical and simulated distributions for the relative frequencies of a given form. Top: Spanish subjunctive case, {\it -se} vs {\it -ra}.  Simulation reported in (A) reproduced the empirical observation of the equivalent distributions (B). Bottom: Intervention of informal institution case, UK vs US variant. Simulations reported in (C) can be compared with the actual empirical distribution (D). Simulated distributions from 200 simulation runs with parameters informed by the fitting procedure and ${\cal W}=2000$.}
\label{sim}
\end{center}
\end{figure}
 
\subsection*{Microscopic dynamics}

As a further assessment, we ran stochastic simulations of the model to reproduce the microscopic evolution of each pair of conventions for the case of spontaneous transition and for the case of the intervention of informal institution. In each numerical experiment we impose the parameters recovered through the fitting procedure described above. We initially consider the case of unregulated (spontaneous) norm adoption. In Fig. \ref{sim}$A$ and $B$ we report probability distribution of observing a relative frequency $n(t)$ for the verbal form {\it -se}, estimated by simulating the evolution of all verbs for which we have empirical record. The simulation results suggest that our model captures well the ensemble evolution over time of the whole empirical distributions. 
Similarly, Fig. \ref{sim}$C$ and $D$ show empirical and numerical results for the class of norm adoption via informal authority, in the case of American spelling change. To account for multiple interventions of informal institutions, numerical experiments were run by `switching on' the parameter $\gamma$ at different, randomly chosen, times. Moreover, the American case consists of conventions that manifest themselves through specific set of words, i.e. the use of {\it or} instead of {\it our} in \textit{behavio(u)r} or \textit{colo(u)r}
or {\it -ize} instead of {\it -ise} in verbs. Thus, for each simulation we extract $\gamma$  from a Gauss distribution centered in $\gamma = 0.02$ (as informed by the data, and $\sigma=0.005$) to reproduce the fact that, in this case, the transition from the old to the new convention is word-dependent. 

By visually comparing empirical distributions of conventions over time for each norm adoption class (inset panel of Fig. \ref{rescaling}, Fig. \ref{sim}$B$, and Fig. \ref{sim}$D$ for formal authority, spontaneous and informal authority respectively) it is evident that, microscopically, the transition from the old to the new convention is governed by different dynamics. For enforcements by formal authorities (inset panel of Fig. \ref{rescaling}), when the norm is regulated the system simply switches to the new convention. On the other hand, for unregulated (spontaneous) norm change (Fig. \ref{sim}$B$) the distribution essentially remains unaltered but for a translation of its mean value which gradually shifts from 1 to 0. Finally, the word dependent transition of the informal institution case yields a broadening of the shape of the distribution over time (Fig. \ref{sim}$D$).

\section*{Conclusion}
In this work we have capitalized on a recently digitized corpus to analyze the process of norm change in the context of the cultural evolution of written English and Spanish. Through the analysis of  $2,541$ cases of convention shifts occurring over the past two centuries, we identified three distinct mechanisms of norm change corresponding to the presence of an authority enforcing the adoption of a new norm, an informal institution recommending the normative update and a bottom-up process by which language speakers select a new norm. Each of these mechanisms displayed different stylized patterns in the data. We rationalized these findings by proposing a simple evolutionary model that describes the actions of the drivers of norm change previously identified in the literature, namely institutions and language users committed to the use of one of the two competing conventions. We showed that this single model captures the dynamics of norm change in each of the three cases described above, quantitatively matching the empirical data in all circumstances. In doing so, it differentiates the empirical curves in three classes according to the measured strength of the institutional intervention (fitted values of $\gamma$ and single curve evaluation, see SI Sec. 8), thus confirming a posteriori the validity of our approach. Finally, through numerical simulations we were also able to reproduce the observed microscopic dynamics of norm adoption.

When a formal institution is present, the transition is sharp and does not depend either on the properties of the considered system (e.g., year or number of published books) or the relative importance of the linguistic convention subject to the norm change. The effect of informal institutions is weaker, resulting in a slower reaction of the system and a smoother transition. Finally, in the bottom-up process of spontaneous change the mechanisms of imitation and reproduction are key in bringing about the relatively slower onset of the new norm, catalyzed by the presence of `committed activists' \cite{vivian1979spelling,scragg1974history,Martin2014}.

It is important to delimit the scope of our findings. First, we only considered cases for which historical records show that a norm change did occur and we did not attempt to predict whether a specific form is at risk of being substituted or not \cite{lieberman2007quantifying}. Second, we considered that the new convention had an advantage over the old one, represented either by the intervention of an institution or by the presence of committed users \cite{vivian1979spelling,scragg1974history,Martin2014}, and we did not consider examples where random drift is the dominant evolutionary force \cite{Plotkin2017Detecting}. In this respect, it is worth mentioning that the role of a committed minority has been investigated in the context of various multi agent models where it has been shown to play an important role on the final consensus provided its size exceeds a certain threshold \cite{xie2011social,mistry2015committed,niu2017impact,baronchelli2018}, as observed also in recent laboratory experiments \cite{ng_cm_science_2018}. Third, we focused on the case where the competition takes place between two alternative norms, but more complex cases where more conventions concur to the process of norm change could exist \cite{gonccalves2017fall}. Fourth, the model we introduced describes the process of norm change for an isolated linguistic group and does not address the important case of language change resulting from the contact between two linguistically independent populations or conflict between different languages \cite{croft2000explaining}.  Finally, our analysis did not consider regional differences or any geographical factors. All these points represent directions for future work.

Taking a broader perspective, our results shed new light on the dynamics leading to the adoption of new linguistic conventions and have implications on the more general process of norm change. 
Today's technology, and in particular online social networks, are reportedly speeding up the process of collective behavioral change \cite{kooti2012emergence,centola2015spontaneous} through the adoption of new norms \cite{ng_cm_science_2018,becker1999constructing,bicchieri1999great,del2016spreading}.
Understanding the microscopic mechanisms driving this process and the signature that it may leave in the data will lead to a better understanding of our society as well as to possible interventions aimed at contrasting undesired effects. In this perspective, we anticipate that our work will be of interest also to researchers investigating the emergence of new political, social, and economic behaviors \cite{nyborg2016social,valenzuela2014facebook}.

\section{Methods}
Google Ngram data set provides about $4\%$ of the total number of books ever printed \cite{michel2011}. We analyzed the following data.
Regulation by a formal institution: 23 Spanish words that change their spelling recovered in \cite{vaquera1986historia,palomo1992,wagner2016,merin2014academia,alcoba2007ortografia,ntlle,casares1954academia} (See SI Sec. 3). 
Intervention of informal institutions: 900 words with the double American and British spelling as reported in SI Sec. 10.A. The list is extracted from \cite{ukuslist} and the double spelling verified with the Merriam-Webster dictionary \cite{webdict}. 
Unregulated evolution: (i) Case of Spanish past subjunctive, 1571 Spanish verbs, 325 of which irregular. 
All the verbs, together with their declination, are listed in \cite{lista_verbi_irr, lista_verbi_reg}. 
(ii) Case of Americanization of English, 46 among words and expressions (the complete list is provided in SI Sec. 10.C). 

For the microscopic dynamics, we performed numerical simulations of the model. At the beginning, we set ${\cal W}_{0}$ conventions in the state ${\cal O}$. At each time authors extract and replace the conventions of the previous time with the following rules. With probability $c$ the author is committed. If the commitments support the new conventions, a convention in the state ${\cal N}$ is added, otherwise a convention in the state ${\cal O}$ is added; with probability $\left(1-c\right)\left(1-\gamma\right)$ the author reproduce the convention extracted; and with probability $\left(1-c\right)\gamma$ the author follows the institution effort: with probability $\left(1-c\right)\gamma E_{\cal N}$ a convention in the state ${\cal N}$ is added while with probability $\left(1-c\right)\gamma E_{\cal O}$ a convention in the state ${\cal O}$ is added. We impose the values recovered by the fitting procedure to set the parameters $c$ and $\gamma$ in the simulations.


\section{Acknowledgements}
We are very grateful to JMR Parrondo and J. Cuesta, who contributed with ideas and discussions to earlier stages of this work.  
RA and AD-G acknowledge support from Ministerio de Economia y Competitividad of Spain project no. FIS2012-38266-C02-02 and no. FIS2015-71582-C2-2-P (MINECO/FEDER) and Generalitat de Catalunya grant no. 2014SGR608. 
LL acknowledges funding from EPSRC Early Career Fellowship EP/P01660X/1.




\newpage

\noindent {\LARGE\textbf{Supporting Information}}
\section*{Symmetric case for eq. (1)}

When the commitment supports the old convention ${\cal O}$, eq. $(1)$ takes the form:
\begin{eqnarray}
n(t+1) &=& \left(1-c\right)\left(1-\gamma\right)n(t)+\left(1-c\right) \gamma E_{{\cal N}} \nonumber \\ 
o(t+1) &=& \left(1-c\right)\left(1-\gamma\right)o(t)+\left(1-c\right) \gamma E_{{\cal O}} +c \,.
\label{connew}
\end{eqnarray}

The general solution of the system of equations \ref{connew} is:

\begin{eqnarray}
n(t) &=& \frac{B}{(1-A)}\left(1-A^{t}\right)+n_{0}A^{t}\nonumber \\
o(t) &=& 1-n(t)\,,
\label{eq_sol_a}
\end{eqnarray}
where $A=\left(1-c\right)\left(1-\gamma\right)$, $B=\left(1-c\right)\gamma E_{{\cal N}}$ and $n_{0} = n(t = 0)$. 

\section*{Spanish past subjunctive}
In Spanish two equivalent forms exist to construct the past subjunctive : the one ending in $-ra$  and the one ending in $-se$ (as in {\it pensa-ra} and {\it pensa-se} ‘had thought’). The form $-se$ evolved from the Latin plusquamperfect subjunctive, while the form $-ra$ evolved from the Latin plusquamperfect indicative \cite{wilson1983}. 

\section*{Spanish spelling reforms}
\label{Spa_regulation}
We present the complete list of the 23 words examined in the Spanish spelling change case, grouped into their respective reforms  \cite{vaquera1986historia,palomo1992,wagner2016,merin2014academia,alcoba2007ortografia,ntlle,casares1954academia}.
\begin{itemize}
\item 1815 :
\begin{itemize}
\item antiquario $\to$ anticuario (antiquarian)
\item quaderno $\to$ cuaderno (notebook)
\item quadro $\to$ cuadro (picture)
\item quando $\to$ cuando (when)
\item quanto $\to$ cuanto (how much)
\item quarto $\to$ cuarto (fourth)
\item quatro $\to$ cuatro (four)
\item quociente $\to$ cociente (quotient)
\item quota $\to$ cuota (quote)
\item quotidiano $\to$ cotidiano (daily)
\item Equador $\to$ Ecuador (Ecuador)
\item iniquo $\to$ inicuo (iniquitous)
\item obliquo $\to$ oblicuo (oblique)
\end{itemize}
\item 1884
\begin{itemize}
\item guion $\to$ gui{\'o}n (script)
\item truhan $\to$ truh{\'a}n (rogue)
\item virey $\to$ virrey (viceroy)
\item vireina $\to$ virreina (viceroy's wife)
\item vireinato $\to$ virreinato (viceroyalty)
\end{itemize}
\item 1911
\begin{itemize}
\item {\'o} $\to$ o (or)
\item {\'a} $\to$ a (to)
\end{itemize}
\end{itemize}
\begin{itemize}
\item 1954
\begin{itemize}
\item di{\'o} $\to$ dio (it gave)
\item fu{\'e} $\to$ fue (it was)
\item vi{\'o} $\to$ vio (it saw)
\end{itemize}
\end{itemize}

\section*{Persistence in time of the old norm}
\begin{figure}[h!]
\begin{center}
\includegraphics[width=0.5\textwidth]{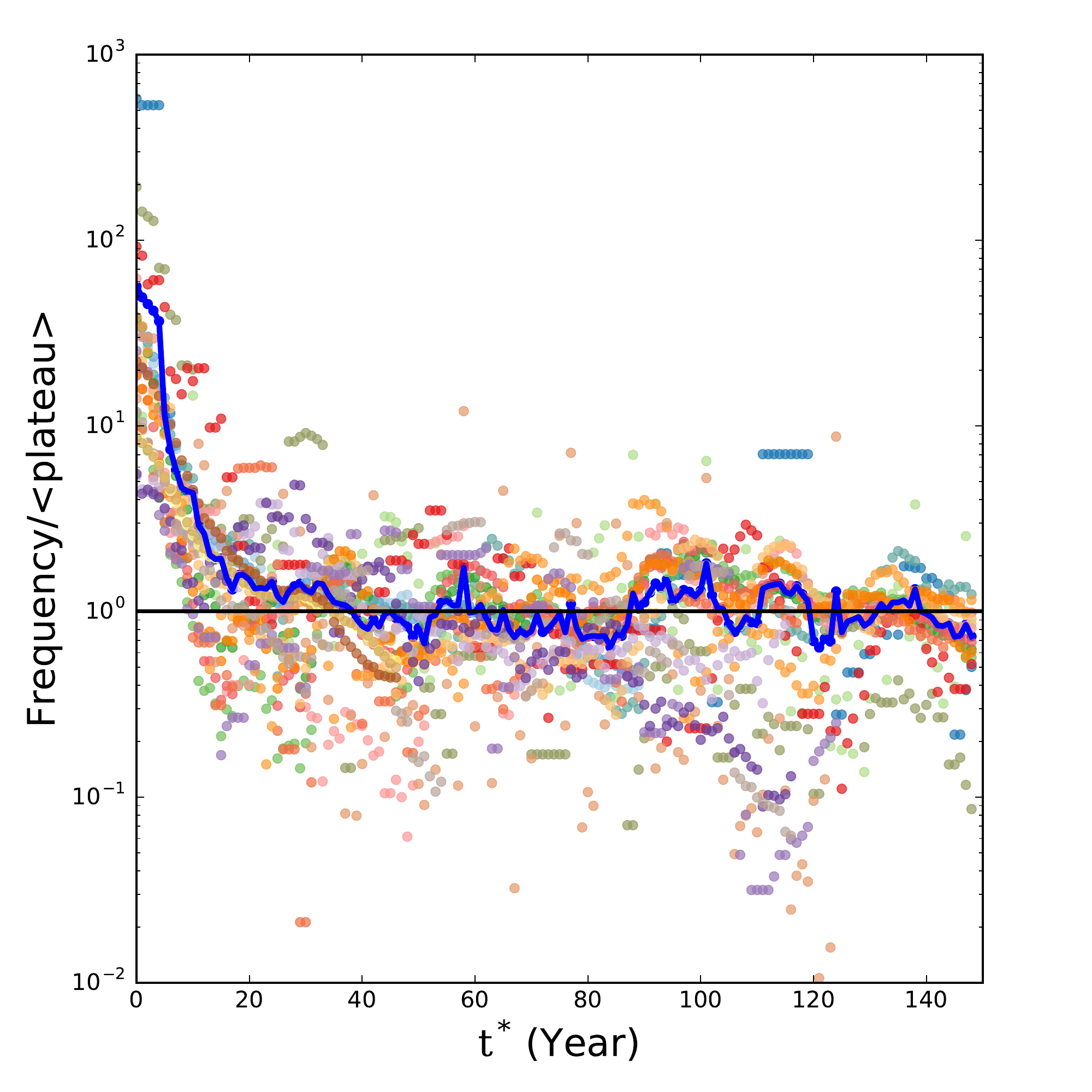}
\caption{Frequency of appearance of the old spelling form as a function of the rescaled time $t*$ for the case of formal regulation. Colored dots represent all the considered words and the blue solid line is the average. The frequency of appearance of each word is divided by the average made on the respective plateau. }
\label{dyn_spa}
\end{center}
\end{figure}

Fig. \ref{dyn_spa} shows that  typically the frequency of appearance of the old spelling in the case of formal regulation does not go to zero but rather, after a sudden drop, stabilizes on a plateau as if a small number of writers ignored the intervention of the institution (i.e., as if they were committed against the new norm). This is mainly due to the presence of historical books, compendiums of the language as well as re-editions of past books (see Supporting Online Material of \cite{michel2011}). For example the list of Spanish books still containing the word "quando" after regulation is composed of texts such as "Biblioteca hist{\'o}rica de la filolog{\'i}a castellana" ("Historical library of Castillan filology"), published in 1893, "Documentos para la historia de la Revoluci{\'o}n de 1809" ("Documents for the history of 1809 revolution") published in 1954 or "Manuscritos lit{\'u}rgicos de las bibliotecas de Espa{\~n}a" ("Liturgical manuscripts of Spanish libraries") published in 1977, which clearly refer to its historical use.

\section*{American Spelling: Important Moments}
\label{US_reg}
Important moments in the history of American vs British spelling:
\begin{itemize}
\item 1806 -  Noah Webster published {\it 'A Compendious Dictionary of the English Language'}
\item 1828 - First American Dictionary {\it 'An American Dictionary of the English Language'}
\item 1848 - Alexander John Ellis published {\it 'A Plea for Phonetic Spelling'}
\item 1876 - American Spelling Reform Association were founded and start to adopt the reforms
\item 1883 - The Chicago Tribune newspaper start to adopt the reforms
\item 1906 - The Simplified Spelling Board was founded and President of the United States Theodore Roosevelt signed an executive order imposing the use of reformed spelling in the official communications of the Congress.
\item 1919 - H.L. Mencken published the first edition of {\it The American Language}
\item 1926 - Henry Fowler published the first edition of {\it Dictionary of Modern English Usage}
\item 1969 -  Harry Lindgren published {\it Spelling Reform: A New Approach}
\end{itemize}

\section*{British and American spelling conflict}

\begin{figure}[t]
\begin{center}
\includegraphics[width=0.5\textwidth]{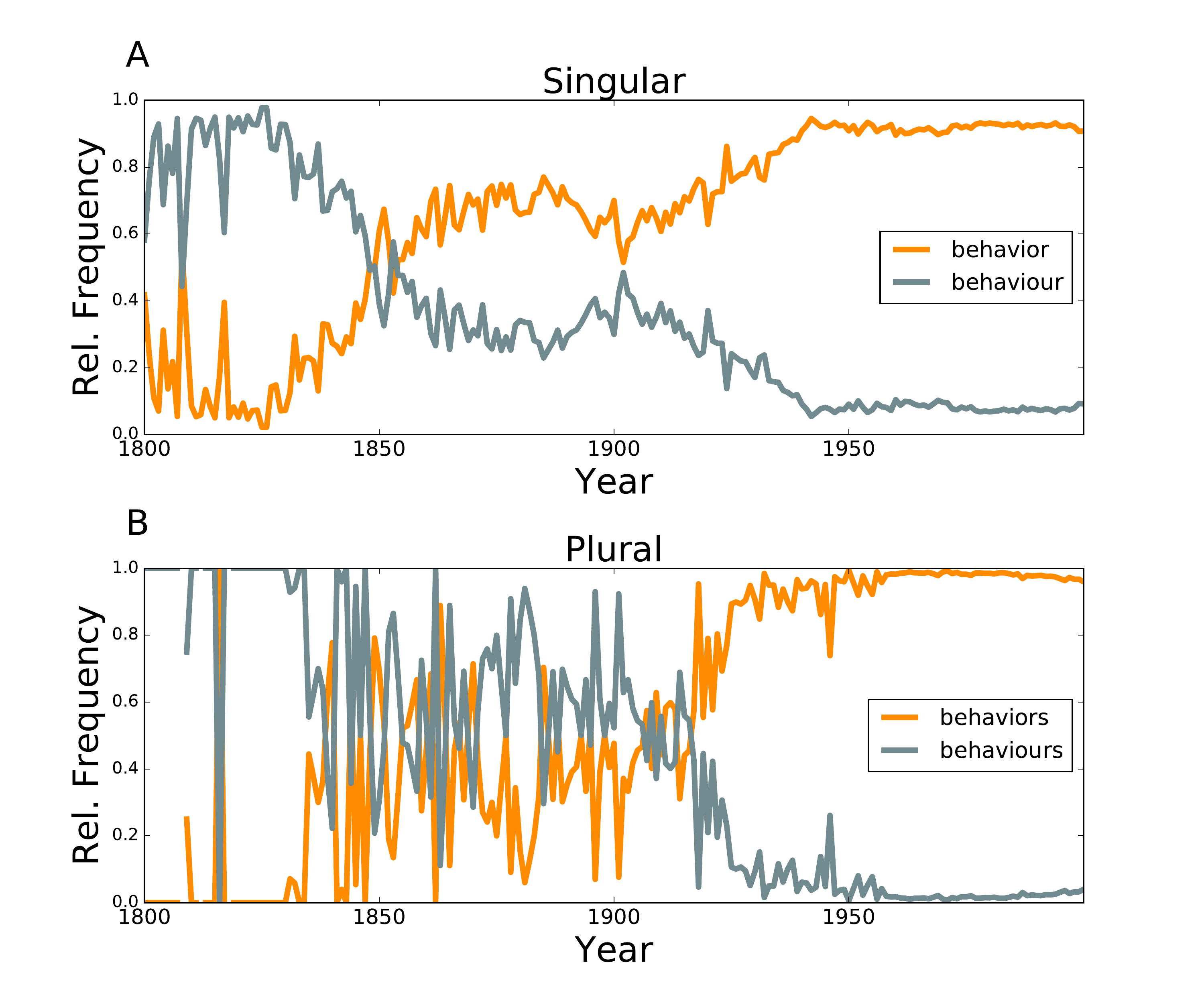}
\caption{(A) Evolution of \textit{behavior}, the American form and \textit{behaviour}, the Bristish form. (B) Evolution of the plural \textit{behaviors} and \textit{behaviours}. }
\label{sup1}
\end{center}
\end{figure}

In our analysis of the case of spelling conflicts, the various inflections of a term, such as singular or plural, or, for a verb, present, past, etc,  were considered separately because they do behave differently. As en example, we report in Fig. \ref{sup1} the evolution of the singular and the plural of the word ``behavior/behavior" (American/British spelling). The American spelling of the singular exceeds that of the British almost a century before the plural one. As mentioned in the Main Text, investigating the reasons of this phenomenon is beyond the scope of our analysis, and an interesting starting point for future work.

\section*{Individual variations beyond the average}

\begin{figure}[!h]
\begin{center}
\includegraphics[width=0.8\textwidth]{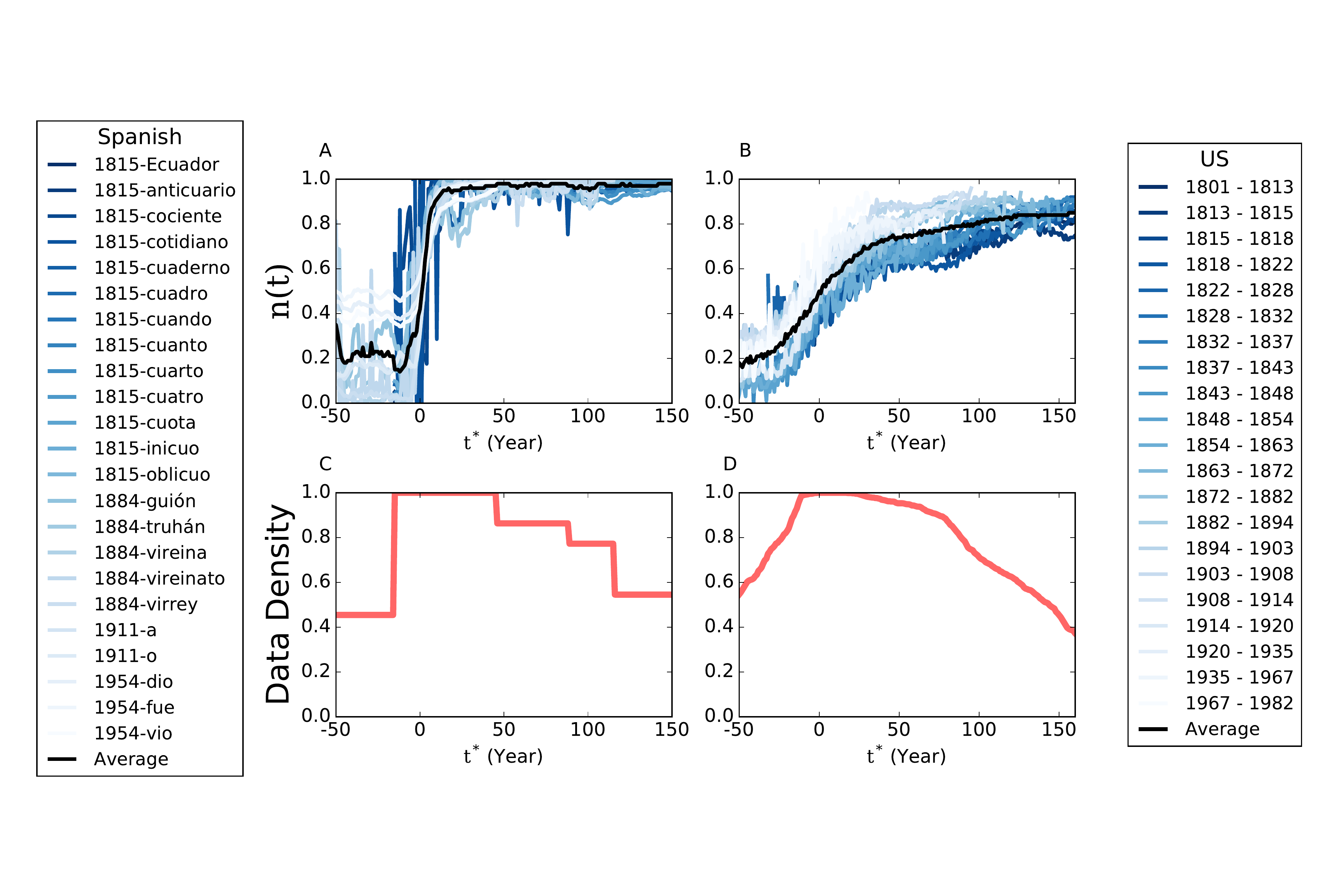}
\caption{$A,C$ Formal Regulation Case. $A$ Relative frequency of the new spelling form as a function of the rescaled time $t*$. Colored lines represent all the considered pairs of words, while the black line is the average. $C$ shows the density of data that contribute to the average for each year. $B,D$ Informal Regulation Case. $B$ Relative frequency of the new spelling form as a function of the rescaled time $t*$. The words are grouped so that each time interval contains $5\%$ of the curves. $D$ shows the density of data that contribute to the average for each year. Colored lines represent the average in these intervals and the black line is the total average. }
\label{contr}
\end{center}
\end{figure}

For the case of formal institution, time is rescaled according to the regulation year (i.e., 1815, 1884, 1911 or 1954, in our dataset, see Sec. \ref{Spa_regulation} above), which in all cases mark the sudden adoption of new norm.
Fig. \ref{contr}$A$, shows how individual pair of norms react to the corresponding regulation event.

For the case of informal regulation, a first dictionary proposing the considered spelling reforms was edited in the 1828 following two decades of discussions (Sec. \ref{US_reg}). The moment in which each American form surpassed in popularity the corresponding British one is distributed over 1800 and 2000. Fig. \ref{contr} $B$ shows how pairs of conventions corresponding to different surpassing times contribute to the global average. For the sake of clarity, individual curves were grouped so that each time window contains $5\%$ of the data. Thus, for example, the curve labeled as  1837-1843 represents the average over the words for which the American spelling surpassed the British one in that time interval.

Figs. \ref{contr} $C$ and $D$ report the density of data contributing to the total average for the corresponding year for the Formal and Informal Regulation cases, respectively.

\section*{Comparison of the three classes}
\begin{figure}[h!]
\begin{center}
\includegraphics[width=0.5\textwidth]{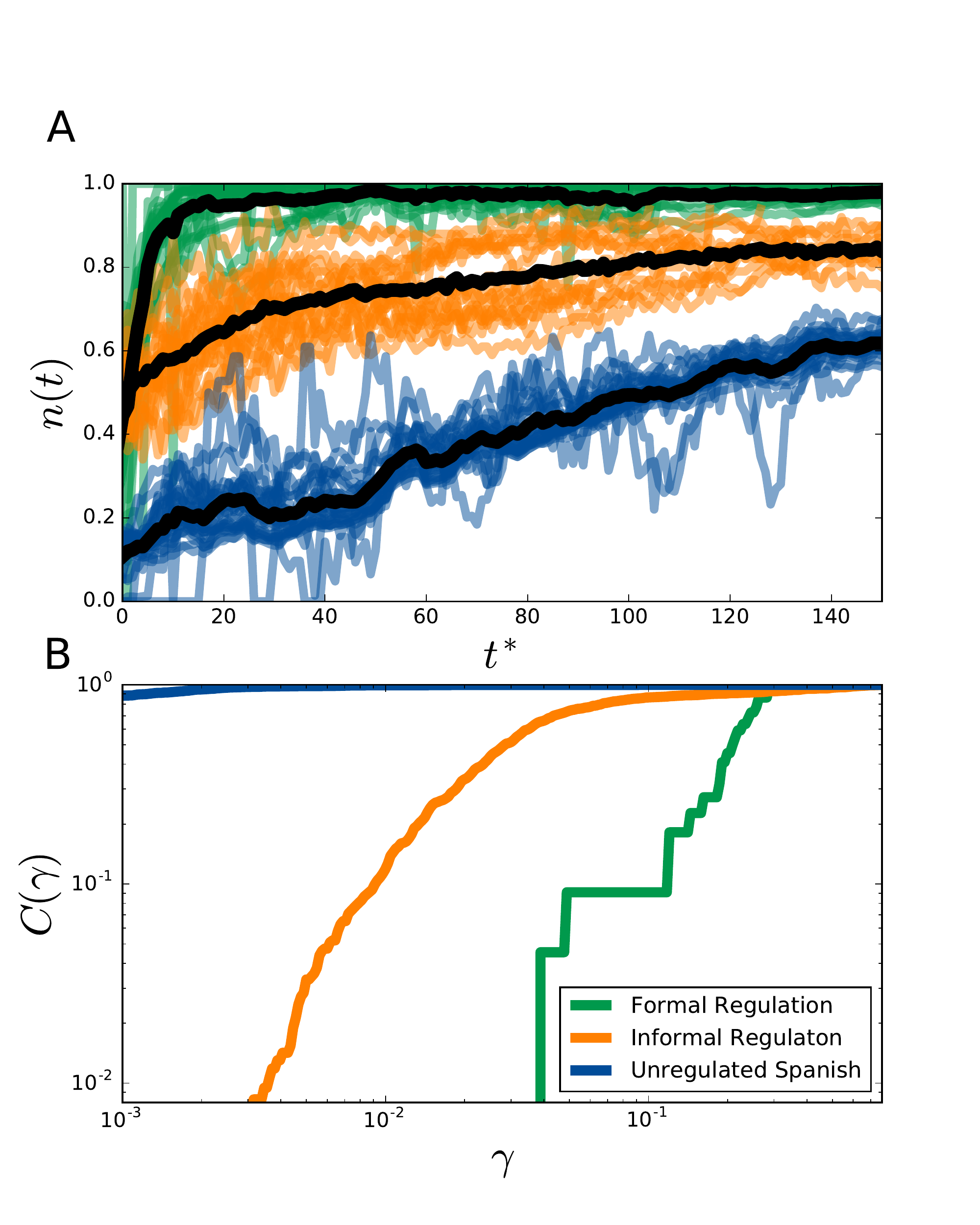}
\caption{$A$ Relative frequency of the new forms as a function of the rescaled time $t^{*} =  t-\tau$. For the case of formal regulation, $\tau$ is the year of regulation, for the case of informal regulation is the year in which the new norm becomes more frequent than the old one, while for the unregulated case $\tau=0$  as the curves are not rescaled. Black lines represent the average. $B$ Cumulative distribution of $\gamma$, $C(\gamma)$,  for each pair of curves for the Formal Regulation, Informal Regulation  and Unregulated (Spanish) cases.  
By comparing the three distributions with the KS test we recover $p=10^{-12}$, $p=10 ^{-20}$ and  $p=0.0$, respectively for the first with the second, the first with the third and the second with the third distribution.  }
\label{gamme}
\end{center}
\end{figure}

To further assess the contribution of the individual pairs of conventions to the three classes we identified in the paper (formal institution, informal institution, and unregulated change),  Fig. \ref{gamme} $A$ shows the temporal evolution of the relative frequency of the new forms,$n(t)$, confirming that the time scales involved in the process of norm change are different for the three cases. 

We further investigated the variability between cases by using the model discussed in the Main Text as an inspecting tool. Fig. \ref{gamme} $B$ reports the cumulative distributions of the parameter $\gamma$, $C(\gamma)$, obtained by fitting individual curves for the three cases.
The Kolmogorov-Smirnov returns $p =10^{-12}$, $p=10 ^{-20}$ and  $p=0.0$, for the first with the second, the first with the third and the second with the third curve respectively. We therefore reject the hypothesis that the three empirical distributions are instances of the same probability distribution.

\section*{Google Books Corpus}

Google Ngram Corpora  offers an unprecedented opportunity to analyze linguistic and cultural change in a quantitative way\cite{michel2011,dodds2011temporal,dodds2015human,koplenig2017impact,montemurro2016coherent,gray2018english}.
The dataset is the product of a massive effort in text
digitization, in collaboration with thousands of the world's libraries \cite{michel2011}. The resulting 5 million books contain over half a trillion words, 361 billion of which are in English \cite{michel2011, pechenick2015characterizing}.

Although Google Ngram can serve as a useful barometer of lexical change \cite{gray2018english} it is important to mention some of its limitations.
To avoid breaking any copyright laws, the datasets are not accompanied by any metadata regarding the texts the corpora consist of \cite{koplenig2017impact} and it does not account for the popularity of a text, namely each text contributes with equal weight to token counts \cite{pechenick2015characterizing,gray2018english}. Furthermore, as recently pointed out, the inclusion of scientific texts, which have become an increasingly substantive portion of the corpus after the 1900s risks to artificially skew the statistical composition of the dataset \cite{pechenick2015characterizing}.

As far as our analysis is concerned, the problem of increasing scientific publications would affect mainly the American English dataset after the 1900s. However, the fact that (i) $71\%$ of the norm-change events we observe in the American English corpus date before 1900, and (ii) we find no significant change in behavior between the transitions observed before and after 1900s or (iii) between the transitions occurring in American English and Spanish (in the case of unregulated change, where we can compare them) are all indications in favor of the robustness of our results. Moreover, it is reasonable to assume that our American English dataset, thanks to its size (946 cases of norm change, in total), includes words whose frequency of use in the scientific literature vary considerably. Thus, the homogeneity of behavior revealed by the curve-by-curve analysis reported in Figs. \ref{contr} and \ref{gamme} further confirms the validity of our findings.


\section*{Dataset composition}
See complete SI on the Publisher's website.

\end{document}